%% file: main.tex
\documentclass[runningheads]{llncs}

\usepackage{graphicx}
\usepackage{amsmath,amssymb}
\usepackage{xspace}
\usepackage{multirow}
\usepackage[labelformat=simple]{subcaption}
	
\usepackage[ruled,linesnumbered]{algorithm2e}
\usepackage{tikz}
	\usetikzlibrary{decorations.pathreplacing,calc}
\usepackage{hyperref}

\usepackage[colorinlistoftodos, shadow]{todonotes} 
\newcounter{todocounter}

\newcommand{\R}{\ensuremath{\mathbb{R}}\xspace}
\newcommand{\Rnonneg}{\ensuremath{\R_{\geq 0}}\xspace}
\newcommand{\A}{\ensuremath{\boldsymbol{A}}\xspace}
\newcommand{\HA}{\ensuremath{\mathcal{H}}\xspace}
\newcommand{\states}{\ensuremath{X}\xspace}
\newcommand{\X}{\ensuremath{\mathcal{X}}\xspace}
\newcommand{\U}{\ensuremath{\mathcal{U}}\xspace}
\newcommand{\V}{\ensuremath{\mathcal{V}}\xspace}
\newcommand{\Tmax}{\ensuremath{T_\mathit{horizon}}\xspace}
\newcommand{\Tsample}{\ensuremath{T_\mathit{sample}}\xspace}
\newcommand{\locs}{\ensuremath{L}\xspace}
\newcommand{\flow}{\ensuremath{F}\xspace}
\newcommand{\inv}{\ensuremath{I}\xspace}
\newcommand{\grd}{\ensuremath{G}\xspace}
\newcommand{\asgnX}{\ensuremath{R_\states}\xspace}
\newcommand{\asgnC}{\ensuremath{R_\ensuremath{C}}\xspace}
\newcommand{\dyn}{\ensuremath{\mathcal{D}}\xspace}
\newcommand{\loc}{\ensuremath{\ell}\xspace}
\newcommand{\traj}{\ensuremath{\mathit{\xi}}\xspace}
\newcommand{\apply}[1]{\ensuremath{\mathit{apply}(#1)}\xspace}
\newcommand{\sstate}[3]{\ensuremath{(#1, #2, #3)}\xspace}
\newcommand{\somesstates}{\ensuremath{\mathcal{R}}\xspace}
\newcommand{\dwell}{\ensuremath{\mathit{Post_C}}\xspace}
\newcommand{\jump}{\ensuremath{\mathit{Post}_D}\xspace}
\newcommand{\tmin}{\ensuremath{t_\mathit{min}}\xspace}
\newcommand{\tshift}{\ensuremath{t_\mathit{shift}}\xspace}
\newcommand{\algo}{\texttt{cont\_reach}\xspace}
\newcommand{\jitter}{\ensuremath{\zeta}\xspace}
\newcommand{\asgn}{\ensuremath{:=}\xspace}

\begin{document}
\title{Efficient reachability analysis of parametric linear hybrid systems with time-triggered transitions\thanks{This research was supported in part by the Austrian Science Fund (FWF) under grant Z211-N23 (Wittgenstein Award) and the European Union's Horizon 2020 research and innovation programme under the Marie Sk{\l}odowska-Curie grant agreement No.\ 754411.}}
\titlerunning{Reachability analysis with time-triggered transitions}
\author{Marcelo Forets\inst{1} \and 
Daniel Freire\inst{2} \and 
Christian Schilling\inst{3}} 
\authorrunning{M.\ Forets et al.}
%
\institute{DMA, CURE, Universidad de la República, Uruguay
\\
\and
Instituto de Física, Facultad de Ciencias, Universidad de la República, Uruguay
\and
IST Austria, Klosterneuburg, Austria
}
\maketitle              

\begin{abstract}
Efficiently handling time-triggered and possibly nondeter\-ministic switches for hybrid systems reachability is a challenging task. In this paper we present an approach based on conservative set-based enclosure of the dynamics that can handle systems with uncertain parameters and inputs, where the uncertainties are bound to given intervals. The method is evaluated on the plant model of an experimental electro-mechanical braking system with periodic controller. In this model, the fast-switching controller dynamics requires simulation time scales of the order of nanoseconds. Accurate set-based computations for relatively large time horizons are known to be expensive. However, by appropriately decoupling the time variable with respect to the spatial variables, and enclosing the uncertain parameters using interval matrix maps acting on zonotopes, we show that the computation time can be lowered to $5{,}000$ times faster with respect to previous works. This is a step forward in formal verification of hybrid systems because reduced run-times allow engineers to introduce more expressiveness in their models with a relatively inexpensive computational cost.

\keywords{Hybrid systems \and Reachability \and Time-triggered transitions}
\end{abstract}
\section{Introduction} \label{sec:intro}

Timed systems play an important role for modeling and analyzing cyber-physical systems.
In this work we consider a class of hybrid automaton models~\cite{AlurCHH92,Henzinger96} with continuous dynamics and discrete events that are time triggered and following a periodic clock.

We propose a reachability framework that rigorously computes an overapproximation of the states reachable by such systems.
Conventional reachability-analysis techniques for hybrid automata tightly integrate the computation of the continuous behavior and the computation of the discrete events.
Our approach allows to separate these concerns, which simplifies the analysis drastically in practice, with respect to both precision and performance.

Our framework is parametric in the analysis tool for the continuous behavior such that we can plug in any reachability algorithm from the literature.
We demonstrate this benefit by using different algorithms for different scenarios.

As a case study, we consider a parametric model of a cyber-physical system consisting of an experimental electro-mechanical brake and a software-implemented periodic controller. The model was originally described in earlier work~\cite{StrathmannO15} where the authors develop a simplified version of a nonlinear system. Although that version of the model is not used in the automotive industry, it is representative of real challenges and allows the application of formal methods for its development.
Computing the reachable states for this simplified model takes twelve hours using a state-of-the-art tool~\cite{StrathmannO15}.
With our approach we are able to analyze the model in less than a minute.

This paper makes the following original contributions:

\begin{enumerate}
    \item We present an efficient algorithm for deterministic periodic time-triggered hybrid systems, where we consider uncertain parameters of the system dynamics or the initial conditions.

    \item We extend the algorithm to the more difficult and new scenario of nondeterministic periodic discrete switches.

    \item We demonstrate the efficiency of our algorithm on a model of an electro-mechanical brake system, which is representative of real challenges in the automotive industry.
\end{enumerate}

\subsection{Related work}

This paper builds on previous works from the reachability literature.
The problem of reachability analysis for purely continuous linear systems with possibly uncertain initial conditions and inputs is essentially solved, since wrapping-free algorithms -- i.e., algorithms that avoid the accumulation of approximation errors -- are known~\cite{GirardGM06} and the sets of reachable states can be computed as closely as desired~\cite{GuernicG10}.
However, hybrid systems with linear continuous dynamics and discrete events are much harder to analyze.
Although these systems have been extensively studied~\cite{KurzhanskiV00,GuernicG09,AlthoffSB10} and efficient implementations are available~\cite{FrehseGDCRLRGDM11,Althoff15}, the discrete events cause inherent issues, mainly because they require conversions from efficient set representations to, e.g., polytopes and back, which is expensive and comes with a loss of precision.

Separation of clock and non-clock variables during the analysis is implemented in HyDra, which is based on the HyPro library~\cite{SchuppAMK17}, but the algorithmic concepts have not been published.
In~\cite{HakimB19} the authors propose a custom algorithm to analyze hybrid automata with \emph{clocked linear dynamics} based on polytopes in vertex representation for the non-clock variables.
This representation has the advantage that clustering (a technique to mitigate complexity explosion in reachability analysis~\cite{FrehseGDCRLRGDM11}) can be applied to the vertices directly. However, vertex representations do not scale well with the system's dimension, in contrast to zonotope or support-function representations~\cite{althoff2016combining}.
Our framework allows to freely choose the set representation.
Moreover, that work does not consider parameter variations while we can easily integrate an existing algorithm from the literature~\cite{AlthoffSB07,Althoff10}.

Time-triggered events occur naturally in engineering contexts.
But even if the events are space-triggered, Bak et al.\ demonstrate that such systems can, under some assumptions, be transformed into a new hybrid system with uncertain continuous dynamics and time-triggered events~\cite{BakBA17}.

\medskip

The remainder of the paper is organized as follows.
In the next section we introduce notation and common reachability concepts.
In Section~\ref{sec:model} we describe the model of the electro-mechanical brake to illustrate the concepts.
In Section~\ref{sec:analysis} we present our algorithm for deterministic time-triggered hybrid systems and later extend it to the nondeterministic case.
In Section~\ref{sec:results} we report on our experimental evaluation on the brake model.
We conclude in Section~\ref{sec:conclusion}.

\section{Preliminaries} \label{sec:preliminaries}

We consider hybrid systems of mixed continuous and discrete dynamics modeled as hybrid automata~\cite{Henzinger96}.
In this paper we are interested in the interaction of a periodic \emph{clock variable} $T$ that follows the ordinary differential equation (ODE) $\dot{T} = 1$ and more complex \emph{state variables} $x_1, \dots, x_n \in \states$.

The continuous dynamics of the state variables are given as a linear time-invariant (LTI) system of the form
\begin{equation}\label{eq:lti}
	\dot{x}(t) = Ax(t) + Bu(t), \quad u(t) \in \U
\end{equation}
where $A \in \R^{n \times n}$ and $B \in \R^{n \times m}$ are matrices and $\U \subseteq \R^m$ is the input domain of dimension~$m$.
We also consider the case where $A$ is parametric and given as an interval matrix, i.e., the entries of $A$ are intervals; if the differentiation is important, we write interval matrices in bold (e.g., \A).
An LTI system is uniquely characterized by the triple $(A, B, \U)$ and we write $\dyn$ to denote the set of all LTI systems.

The discrete dynamics of our hybrid automata are \emph{time triggered}, by which we mean that events only depend on the values of the clock variable, and \emph{periodic}.
The latter notion is motivated by the common setting in which a digital controller follows a wall clock and sends a periodic signal at multiples of \Tsample seconds.
In practice, not every signal will arrive at exact multiples of \Tsample; instead, some signals may arrive a bit earlier and some may arrive a bit later.
However, if a signal deviates from the exact multiple of \Tsample, say, it arrives at time $\Tsample + \varepsilon$, the following signals will not drift by $\varepsilon$ because the reference is still given by the wall clock.
Finally, updates of state variables upon discrete events are restricted to affine transformations.

Formally, we define the syntax of hybrid automata below (following~\cite{HakimB19}).

\begin{definition}[Hybrid automaton]
	A \emph{hybrid automaton} is a tuple $\HA = (\states, T, \locs, \flow, \inv, \grd, \asgnX, \asgnC)$ where
	\begin{itemize}
		\item $\states = \{x_1, \dots, x_n\}$ is the finite set of state variables,

		\item $T$ is the clock variable,

		\item $\locs = \{\loc_1, \dots, \loc_{|\locs|}\}$ is the finite set of locations,

		\item $\flow : \locs \to \dyn$ assigns continuous affine dynamics to each location,

		\item $\inv : \locs \to 2^{\R}$ assigns an invariant to each location,

		\item $\grd : \locs \times \locs \to 2^{\R}$ assigns a guard to each pair of locations, and

		\item $\asgnX : \locs \times \locs \to \R^{n \times n} \times \R^n$ and $\asgnC : \locs \times \locs \to \R \times \R$ assign affine reset maps for the state and clock variables to each pair of locations.
	\end{itemize}
\end{definition}

We restrict the invariant and guard constraints to convex polyhedra, i.e., intersections of linear constraints $\{t \mid a t \leq b\}$ where $a, b \in \R$.
We call a pair of locations $(\loc_1, \loc_2)$ a \emph{transition} if $\grd(\loc_1, \loc_2) \neq \emptyset$.
Formally, reset maps are given as a matrix and a vector, but we use a common abbreviated notation as a sum of variables and constants (and for variables that are not mentioned, the map is the identity).
In Figure~\ref{fig:hybridAutomatonPIcontroller} we depict the hybrid automaton of an electro-mechanical brake, which will serve as our evaluation model later.

We briefly recall the semantics of hybrid automata; for details we refer to the literature~\cite{Henzinger96}.
We start with the continuous behavior.
The dynamics of the clock variable is trivial.
In a location with LTI system $(A, B, \U)$, given an initial state $x_0 \in \R^n$ and an input signal $u$ such that $u(t) \in \U$ for all $t$, a \emph{trajectory} of~\eqref{eq:lti} is the unique solution $\traj_{x_0,u}: \Rnonneg \rightarrow \R^n$ with
\begin{equation*}\label{eq:pwa_analytic}
	\traj_{x_0,u}(t) = e^{At} x_0 + \int_0^t e^{A(t-s)} B u(s) \, ds.
\end{equation*}

Given a set $\X_0 \subseteq \R^n$ of initial states, an initial clock value $T_0 \in \R$, and an invariant \inv, the \emph{continuous-post operator} \dwell computes the set of reachable state and clock variables at time point $t \in \Rnonneg$ for all input signals $u$ over \U:
\begin{align*}
	\dwell((A, B, \U)&, \X_0, T_0, \inv, t) := \\
	& \{ (\traj_{x_0, u}(t), T_0 + t) \mid T_0 + t \in \inv, x_0 \in \X_0, u(s) \in \U \text{ for all } s \}
\end{align*}
We overload the operator for quantifying over all time points:
\begin{equation*}
	\dwell((A, B, \U), \X_0, T_0, \inv) := \bigcup_{t \in \Rnonneg} \dwell((A, B, \U), \X_0, T_0, \inv, t)
\end{equation*}

The set computed by the \dwell operator is called a \emph{flowpipe}.
We denote the application $A \X + b = \{Ax + b \mid x \in \X\}$ of an affine transformation $M = (A, b)$ to a set of states $\X$ by $\apply{M, \X}$ (and similarly we write $\apply{M, T}$ for a time point $T$).
A \emph{symbolic state} is a triple $\sstate{\loc}{\X}{T} \in \locs \times 2^{\R^n} \times \R$.
The \emph{discrete-post operator} \jump maps a symbolic state to a set of symbolic states by following the outgoing discrete transitions:
\begin{align*}
	\jump\sstate{\loc}{\X}{T} := \hspace*{-1cm} \bigcup\limits_{T \in \grd(\loc, \loc') \land \apply{\asgnC(\loc, \loc'), T} \in \inv(\loc')} \hspace*{-1cm} \{ \sstate{\loc'}{\apply{\asgnX(\loc, \loc'), \X}}{\apply{\asgnC(\loc, \loc'), T}} \}
\end{align*}

The \emph{reach set} of \HA from a set of initial symbolic states $\somesstates_0$ is the smallest set \somesstates of symbolic states such that
\begin{equation*}
	\somesstates_0 \cup \bigcup\limits_{\sstate{\loc}{\X}{T} \in \somesstates} ~ \bigcup\limits_{(\X', T') \in \dwell(\flow(\loc), \X, T, \inv(\loc))} \jump\sstate{\loc}{\X'}{T'} \subseteq \somesstates
\end{equation*}
holds.
The reach set can also be seen as a union of flowpipes.

\medskip

To represent sets of points in Euclidean space, we mainly consider shapes called \emph{zonotopes} in this work.
Zonotopes are specific centrally-symmetric convex polytopes defined as the image of a hypercube under an affine projection, or equivalently as the Minkowski sum of a finite set of line segments~\cite{Ziegler12}.
A zonotope is commonly represented as a center plus a finite set of generators.
The order of a zonotope is the ratio of the number of generators and the dimension.
Zonotopes are closed under Minkowski sums and affine transformations.

\section{A model of an electro-mechanical brake}\label{sec:model}

We consider a model of an electro-mechanical brake with periodic controller, originally proposed by Strathmann and Oehlerking~\cite{StrathmannO15}. In this section we present a high-level description of the hybrid model and describe the dynamic equations. We refer to the original work for further details.

The system consists of an electrical engine that pushes the inner side of a brake caliper. Once the caliper position $x$ reaches a certain threshold $x_0$, the brake disk gets in contact with the wheel and deceleration starts. The pressure between the disk and the wheel (and hence the deceleration) increases the more the caliper is moved beyond the threshold.

The model has several parameters that can be divided into two groups: the physical parameters of the brake hardware and the parameters of the PI controller. The hardware parameters are subject to device erosion and production tolerances. The controller parameters are often modified after deployment to satisfy requirements such as vibrations or noise. Hence we are interested in guaranteeing certain properties for given parameter \emph{ranges}.

The original controller is modeled as a hybrid automaton with three locations: the idle location, where the caliper is positioned at the farthest point from the disk; the positioning location, where the caliper is moved but there is still no contact with the disk; and the force-control location, where the disk is pushed against the wheel and the brake decelerates. The control strategy consists of a model-based feed-forward controller and feedback through a discrete-time PI controller, to account for disturbances.

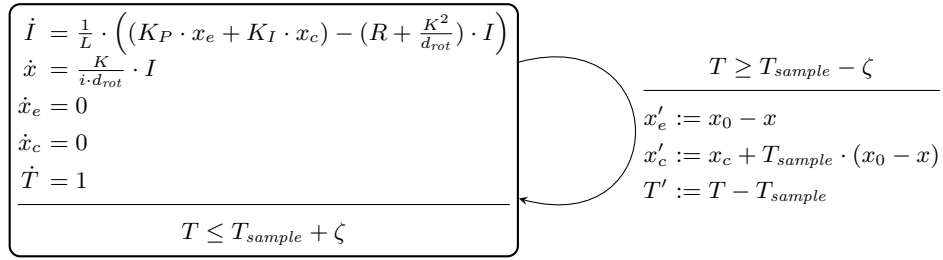
\begin{figure}[t]
	\centering
	\input{embrake_automaton}
	\caption{Hybrid automaton of the electro-mechanical brake plant with periodic discrete-time PI controller and sampling jitter.}
	\label{fig:hybridAutomatonPIcontroller}
\end{figure}

In this work we consider the discrete-time PI controller, which is modeled as a hybrid automaton with a single location and a self-loop that samples the distance $x_0 - x$ at discrete points in time as multiples of the sampling time $\Tsample$. In addition, sampling jitter is taken into account, enabling the discrete transition in a nondeterministic interval $\left[-\jitter, +\jitter \right]$, with $0 \leq \jitter \ll \Tsample$. 
The plant with discrete controller is shown in Figure~\ref{fig:hybridAutomatonPIcontroller}, where the state variable $I$ models the current of the DC motor and other physical measures are assumed constants: $R$ (electrical resistance), $L$ (inductance), $K$ (which accounts for the losses due to the rotational motion), $i$ (transmission ratio of the gearbox), $d_\mathit{rot}$ (friction coefficient for the rotational motion), and $K_P$ and $K_I$ (PI controller constants with inputs $x_e$ and $x_c$, respectively).

We note that more physically accurate, non-linear models exist; however, only the linearized model has been analyzed in~\cite{StrathmannO15}. Parametric reachability for such models poses actual challenges to state-of-the-art verification tools.



\section{Reachability analysis with time-triggered events}\label{sec:analysis}

\begin{figure}[tb]
	\centering
	\input{timeline}
	\caption{Timeline of relevant events in the continuous-time reach-set computation for a location with a single outgoing transition and discretization time step~$\delta$.
	The marked time points are $t_0$ (the starting point), $t_1$ (the first time point when the guard is enabled), $t_2$ (the last time point when the guard is enabled), and $t_3$ (the last time point when the invariant is enabled)}
	\label{fig:timeline}
\end{figure}
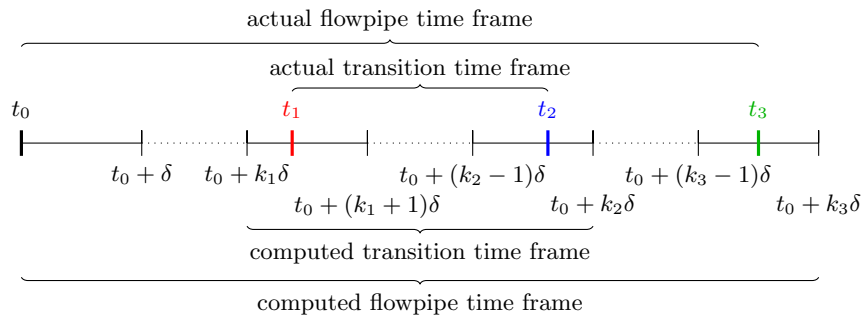

Since precisely computing a flowpipe for the class of systems we consider is not possible, the goal is to compute an overapproximation instead.
Our hybrid automaton model only contains time-triggered transitions: the invariants and guards only depend on the clock variable.
Since the clock variable evolves independently from the other state variables, we can reason about the time frame when the transitions are enabled in isolation.
To simplify the discussion, we concentrate on the clock variable for now and assume that we start at time point~$t_0$ from an initial clock value $T_0 \in \inv(\loc)$ in a location \loc with one outgoing transition $e = (\loc, \loc')$.
The following explanations are illustrated in Figure~\ref{fig:timeline}.

In this work we assume that time is bounded by some constant $\Tmax$.
Hence, considering the continuous flowpipe construction (via the \dwell operator), there is either a latest time point $t_3$ when the \dwell operator returns a nonempty set, namely when the invariant $\inv(\loc)$ is enabled for the last time ($T_0 + t_3 \in \inv(\loc)$ but $T_0 + t_3 + \varepsilon \notin \inv(\loc)$ for any $\varepsilon > 0$), or otherwise we define $t_3 := \Tmax$.
Computation of the flowpipe can be stopped after time point~$t_3$.

Moreover, if the guard $\grd(e)$ is ever enabled, then the points in time where it is enabled form an interval $[t_1, t_2]$ with $t_0 \leq t_1 \leq t_2 \leq t_3$.
Note that, since we do not require that guards are bounded, it is possible that $t_1 = t_0$ and/or $t_2 = t_3$ hold.
We are particularly interested in the flowpipe at the time interval $[t_1, t_2]$ when the transition is enabled because this part will be the input to the discrete-post operator \jump later.

\subsection{Time discretization}

The algorithms that we will use to compute an overapproximation of the flowpipe work by fixed-step time discretization.
This means that they choose a sampling time step $\delta \in \Rnonneg$ and compute an overapproximation for successive time intervals $[t_0, t_0 + \delta]$, $[t_0 + \delta, t_0 + 2 \delta]$, etc.
Thus the flowpipes we deal with in practice are unions of sets, one for each time frame of size $\delta$.
In Figure~\ref{fig:timeline} we depict these time frames together with the important time points identified above.
As can be seen, from those time points we can easily identify the set indices $k_1$, $k_2$, $k_3$ such that we need to compute the sets in the time intervals from $t_0$ to $t_0 + k_3 \delta$ and take the transition for the sets in the time intervals from $t_0 + k_1 \delta$ to $t_0 + k_2 \delta$.

Having determined the relevant time points for the clock variable, we now consider the flowpipe for the state variables (note that computing the flowpipe for the clock variable is trivial).
Depending on the system dynamics, there are several options how to compute a discretization in the literature (e.g.,~\cite{GirardGM06,AlthoffSB07,FrehseGDCRLRGDM11}), but the details are not important here since the general concept is the same:
Given an LTI system $(A, B, \U)$, a set of initial states $\X_0 \subseteq \R^n$, and a step size $\delta \in \Rnonneg$, we compute the matrix $\Phi := e^{A \delta}$ and two sets $\X(0)$ and $\V$.
The set $\X(0)$ overapproximates the flowpipe for time interval $[0, \delta]$; in other words, using the initial clock value $T_0 = 0$ and the invariant $I$ with the linear constraint $T \leq \delta$, we have that $\X(0) \supseteq \dwell((A, B, \U), \X_0, T_0, I)$.
The set $\V$ overapproximates the effect of the nondeterministic inputs such that the set recurrence
\[
	\X(k) = \Phi \X(k-1) \oplus \V, \quad k > 0
\]
satisfies the property that $\X(k)$ overapproximates the flowpipe for time interval $[k \delta, (k+1) \delta]$.
Here $\oplus$ denotes the Minkowski sum.

From the previous analysis we can conclude that we need to compute the sets $\X(k)$ for $k = 0, \dots, k_3 - 1$ and that the transition can only be taken from the sets $\X(k')$ for $k' = k_1, \dots, k_2 - 1$ (recall Figure~\ref{fig:timeline}).

\subsection{Deterministic switches} \label{ssec:deterministic_switches}

\begin{figure}[tb]
	\centering
	\begin{subfigure}[t]{0.48\textwidth}
		\raisebox{11mm}{\input{simple_automaton}}
		\caption{Hybrid automaton.}
		\label{fig:automaton_simple}
	\end{subfigure}
	\hfill
	\begin{subfigure}[t]{0.48\textwidth}
		\includegraphics[width=6cm,height=4cm,keepaspectratio]{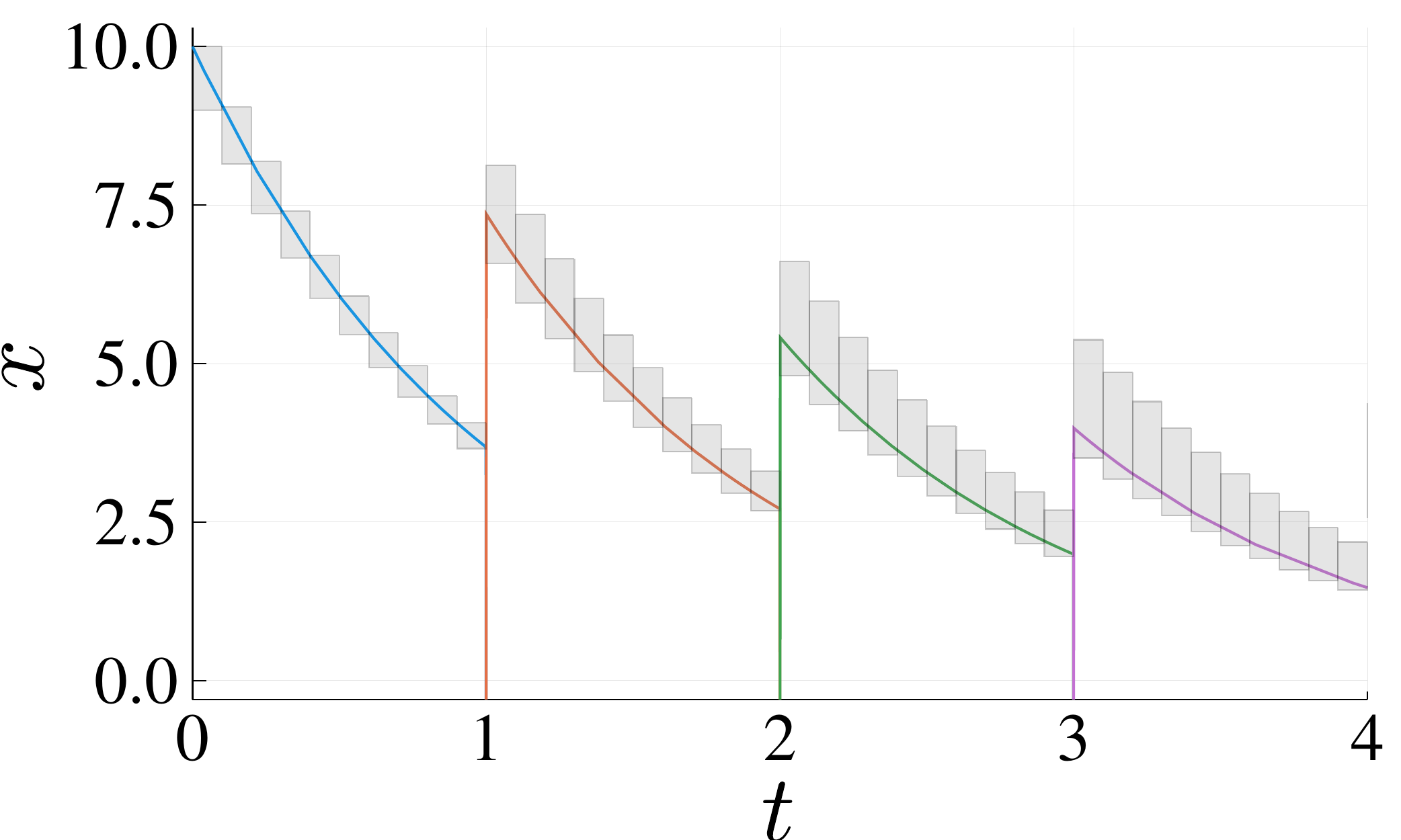}
		\caption{Flowpipes for switches at multiples of the sampling time ($\delta = 0.1$).}
		\label{fig:simple_example_det_sets_multiple}
	\end{subfigure}
	\\
	\begin{subfigure}[b]{0.48\textwidth}
		\includegraphics[width=6cm,height=4cm,keepaspectratio]{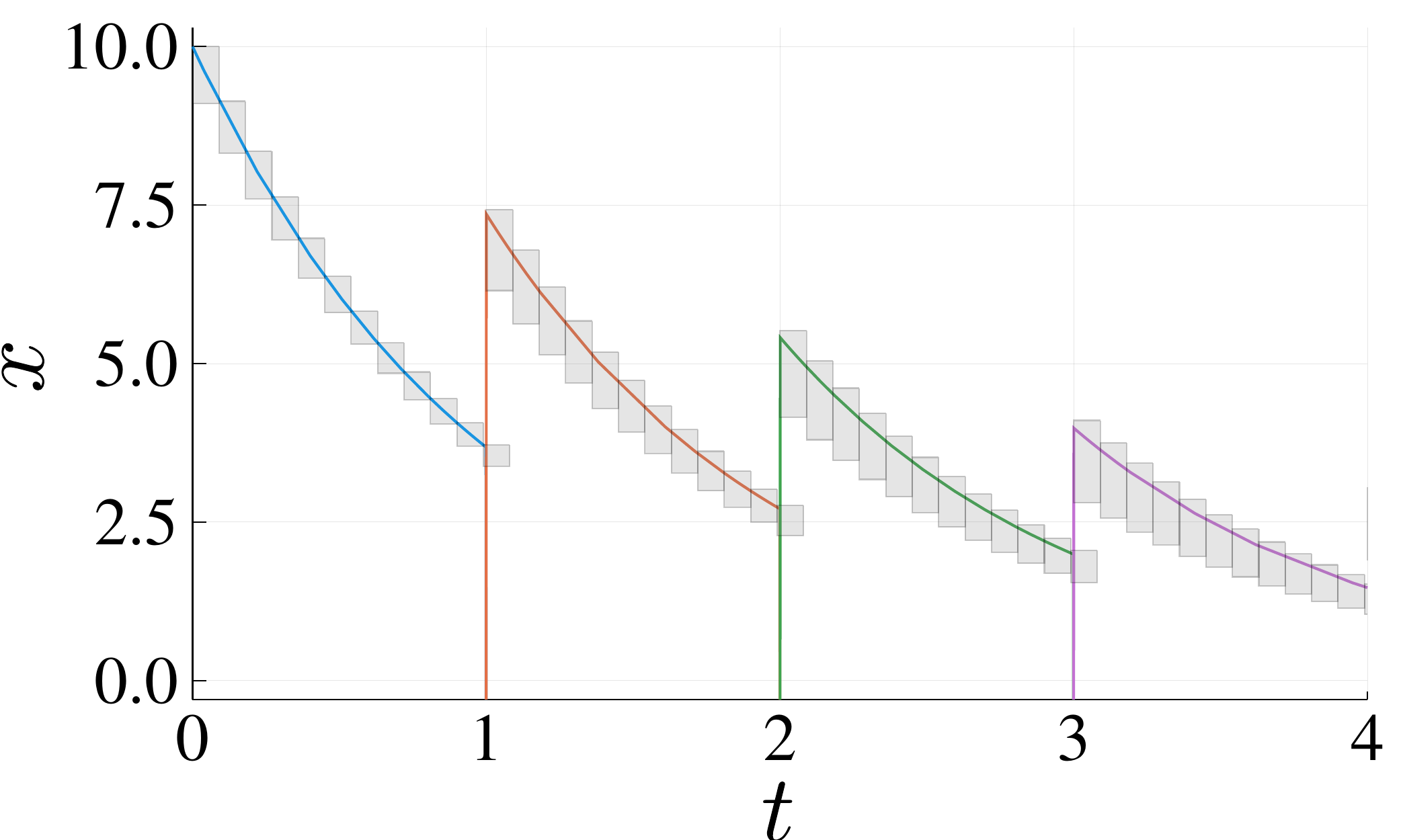}
		\caption{Flowpipes for switches not at multiples of the sampling time ($\delta = 0.09$).}
		\label{fig:simple_example_det_sets_nonmultiple}
	\end{subfigure}
	\hfill
	\begin{subfigure}[b]{0.48\textwidth}
		\includegraphics[width=6cm,height=4cm,keepaspectratio]{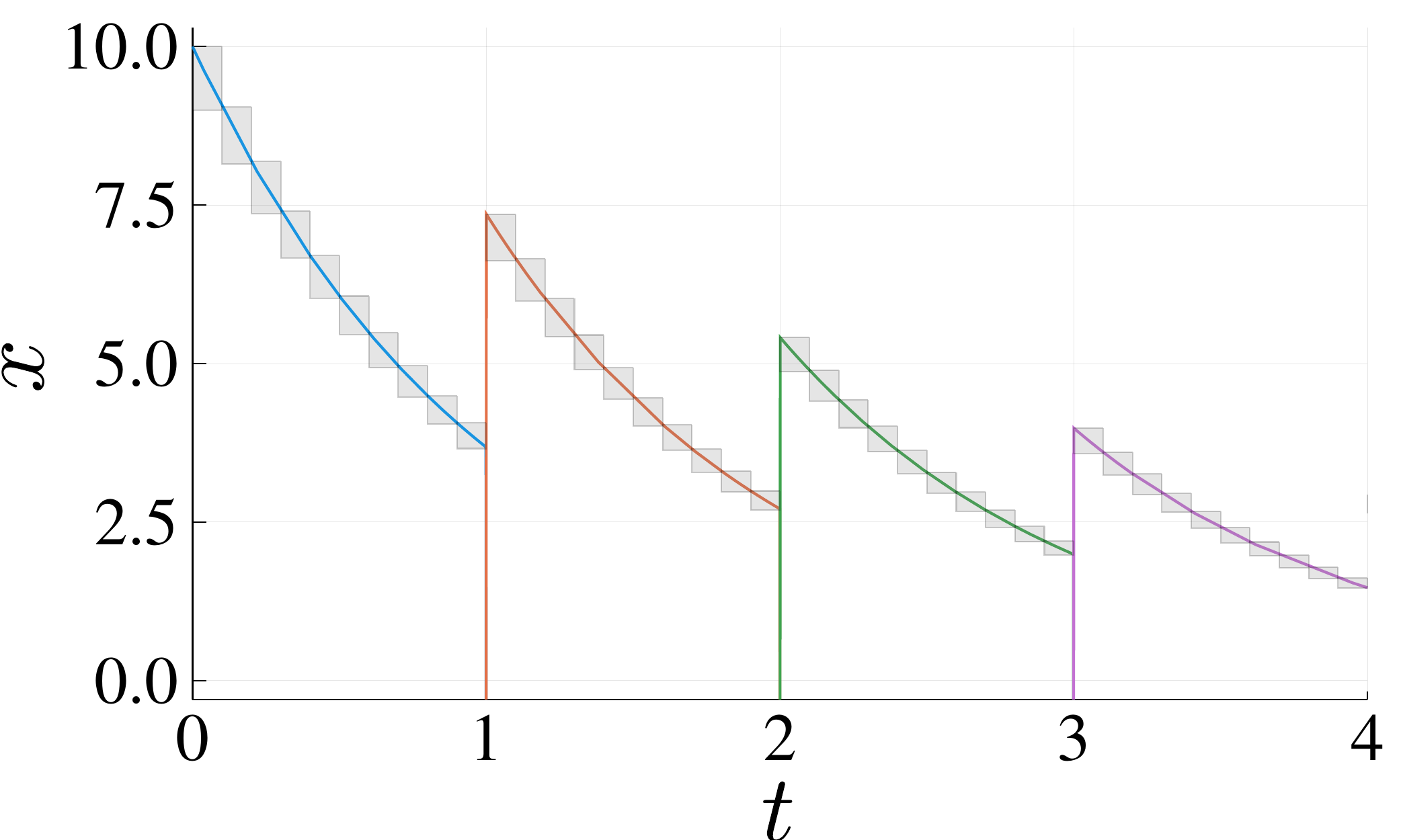}
		\caption{Flowpipes for the setting in Figure~\ref{fig:simple_example_det_sets_multiple} but as described in Section~\ref{ssec:deterministic_switches_improved}.}
		\label{fig:simple_example_det_sets_multiple_precise}
	\end{subfigure}
	\caption{A simple hybrid automaton and trajectories for the deterministic instantiation $\Tsample = 1$ and $\jitter = 0$.
	(The colored vertical lines are plotting artifacts.)}
	\label{fig:simple_example_deterministic}
\end{figure}

We illustrate the above observations on a simple running example.
Consider the hybrid automaton in Figure~\ref{fig:automaton_simple} where for now we consider the case $\jitter = 0$ such that transitions happen at deterministic time points (namely multiples of $\Tsample$).
The other plots in Figure~\ref{fig:simple_example_deterministic} show the (unique) trajectory starting from the initial state $x(0) = 10, T(0) = 0$.
Using an algorithm that will be explained later, we obtain the flowpipes in Figure~\ref{fig:simple_example_det_sets_multiple} covering the trajectory with a time step of~$\delta = 0.1$.
We chose such a large time step for better visualization; the precision is not representative of what can be achieved.

We compute the flowpipes using an algorithm first presented by Girard et al.~\cite{GirardGM06} and generalized in~\cite[Algorithm~4.2]{LeGuernic09}, which we call the \texttt{GLGM06} algorithm.
The algorithm uses zonotopes as set representation.
By a suitable reordering of the set recurrence, the algorithm avoids the wrapping effect of approximation errors, which results in a very precise flowpipe.

\begin{algorithm}[t]\footnotesize
	\caption{Periodic time-triggered flowpipe construction.}
	\label{alg:reach}
	\KwIn{\\
	$(A, B, \U), \X_0$: LTI system and set of initial states \\
	$\Tsample, \jitter$: encoding of the time interval when the transition is enabled \\
	$R$: reset map of the transition \\
	\algo, $\delta$: \dwell approximation algorithm and discretization time step \\
	$\Tmax$: time horizon
	}
	\vspace*{2mm}
	$k_1, k_2$ \asgn compute\_transition\_indices($\Tsample, \jitter, \delta$)\;
	$\X(\cdot)$ \asgn $\algo((A, B, \U), \X_0, \delta, k_2)$\tcp*{compute first flowpipe}
	flowpipes \asgn [$\X(\cdot)$]\;
	$\tmin$ \asgn $k_1 \delta$\;
	$\tshift$ \asgn $\Tsample - \jitter$\;
	$k_2$ \asgn $k_2 + \lceil \frac{\jitter}{\delta} \rceil$\label{line:correction}\tcp*{correction for nondeterministic transitions}
	\While{$\tmin \leq \Tmax$}{
		\tcp{extract solution sets for $[k_1, k_2)$, cluster, apply reset map}
		$\X_1$ \asgn jump($\X(\cdot), k_1, k_2, R$)\label{line:jump}\;
		\BlankLine
		$\X(\cdot)$ \asgn $\algo((A, B, \U), \X_1, \delta, k_2)$\tcp*{compute next flowpipe}
		flowpipes \asgn flowpipes $\cdot$ [shift($\X(\cdot), \tshift$)]\label{line:shift}\tcp*{shift flowpipe by $\tshift$}
		$\tmin$ \asgn $\tmin + k_1 \delta$\;
		$\tshift$ \asgn $\tshift + \Tsample$\;
	}
	\Return flowpipes\;
\end{algorithm}

In Algorithm~\ref{alg:reach} we describe a reachability algorithm based on the abstract discretization scheme.
For simplifying the presentation, in that algorithm we assume that time starts at $t_0 = 0$, the automaton has a single location with a self-loop transition, and that the invariant is violated as soon as the guard gets disabled (i.e., $t_3 = t_2$ in Figure~\ref{fig:timeline}); the generalization to multiple locations and transitions and to an extended invariant is straightforward but requires more boilerplate code.
(Note that this setting applies to the electro-mechanical brake model.)
The inputs to the algorithm are the system information encoded in the hybrid automaton (LTI system, switching time interval, and reset map), the set of initial states $\X_0$, an algorithm \algo to approximate \dwell together with a discretization step size $\delta$, and a time horizon \Tmax.

We first determine the indices $k_1$ and $k_2$ as described previously.
Then we compute the first flowpipe using \algo, starting from $\X_0$.
As a technical detail, we need to adapt the parameter $k_2$ for further flowpipe computations in the nondeterministic case in line~\ref{line:correction}.
The reason is that the very first flowpipe starts from a deterministic time point $t_0$ while the other flowpipes need to account for the possible deviation from multiples of \Tsample.
The while loop then interleaves the computation of a discrete transition and a new flowpipe.

The discrete transition is computed as follows (function ``jump'' in line~\ref{line:jump}):
First we extract the sets $\X(k_1), \dots, \X(k_2 - 1)$ from the previous flowpipe.
Next we apply an operation that is known as \emph{clustering} to these sets; since in the deterministic case we only ever deal with a single set (for which clustering is just the identity), we defer the explanation of clustering to Section~\ref{ssec:nondeterministic_switches}.
Finally, we apply the reset map of the transition to obtain the discrete successors.

From the set of discrete successors, we then spawn the next flowpipe.
We compute this flowpipe as if the reference point was $t_0 = 0$.
To account for that, we shift the resulting flowpipe by the respective amount of time in line~\ref{line:shift}.

\subsection{Improving precision for nonparametric LTI systems} \label{ssec:deterministic_switches_improved}

The precision of the flowpipe construction is inherently limited by the precision of the discretization, which itself depends on the set $\X_0$ of states from which the flowpipe emerges.
Thus it is generally desirable to compute the sets from which a transition is taken with improved precision (the set $\X_1$ in Algorithm~\ref{alg:reach}).
If there is an approximation error in these sets, we propagate this error to the next flowpipe:
In Figure~\ref{fig:simple_example_det_sets_multiple}, the sets in the flowpipes become bigger with each transition.
Figure~\ref{fig:simple_example_det_sets_nonmultiple} shows a similar setting where \Tsample is not a multiple of $\delta$, with similar results.
We observe that the flowpipes are actually tighter, which is because we use a smaller (more precise) time step and the analysis has to compute one more set, which in this case (the interval for $x$) is tighter.

Consider again the case that the transition is taken at a multiple of the discretization step $\delta$, i.e., at time $k \delta$ for some $k > 0$.
Since the set $\X(k-1)$ overapproximates the flowpipe for time interval $[(k-1) \delta, k \delta]$ and the set $\X(k)$ overapproximates the flowpipe for time interval $[k \delta, (k+1) \delta]$, we can conclude that the intersection $\X(k-1) \cap \X(k)$ overapproximates the flowpipe at time \emph{point} $k \delta$.
However, in practice, intersections are expensive to compute and often need to be overapproximated in order to obtain a set representation of the supported class (e.g., a zonotope).
Thus computing the intersection may actually end up being less precise than just taking one of the sets $\X(k-1)$ or $\X(k)$.

For LTI systems without parameter variation, there is actually a better way to obtain a more precise overapproximation of the flowpipe at time point $k \delta$.
Let us denote the exact set of those states as $\X_{k \delta}$.
Then we have that
\[
	\X_{k \delta} \subseteq e^{A k \delta} \X_0 \oplus \bigoplus_{i=1}^k \Phi^{i-1} \V.
\]
Using an appropriate set representation (e.g., zonotopes), this overapproximation can actually be represented exactly (assuming that we can compute the matrix exponential exactly).
The overapproximation is indeed equivalent to the true solution for homogeneous systems.
(For inhomogeneous systems the overapproximation is not exact because \V contains an overapproximation.)

Recall again the simple system from Figure~\ref{fig:automaton_simple}.
The analytic solution of the continuous dynamics at time $t$, starting from a set $x(0) = \X_0$, is $x(t) = e^{-t} \X_0$.
Generalizing to the hybrid system, the analytic solution after taking $k$ transitions at time $t \in (k \Tsample, (k+1) \Tsample)$ is
\begin{equation*}
	\X(t) = 2^{k-1} e^{-(k-1) \Tsample} \X_0 e^{-(t - (k-1) \Tsample)}.
\end{equation*}
Figure~\ref{fig:simple_example_det_sets_multiple_precise} clearly shows improvements in precision over the more general algorithm in Figure~\ref{fig:simple_example_det_sets_multiple}.
In particular, no errors are propagated to the next flowpipe.

\subsection{Nondeterministic switches}\label{ssec:nondeterministic_switches}

We now turn to the case where the discrete transition is nondeterministic.
More precisely, the guard of the transition is enabled at time points from a proper time interval $\Tsample \pm \jitter$.
Algorithm~\ref{alg:reach} already covers this more general case, so we just describe the two main technical differences to the deterministic case.

The first difference is that we need to increment the value $k_2$ by $\lceil \frac{\jitter}{\delta}\rceil$ (line~\ref{line:correction}). This integer corresponds to the extra number of sets that are required from the second flowpipe onward, to account for the asymmetry of the start time $t_0 = 0$, with respect to the transition times, enabled at $k\Tsample - \jitter$ for each $1 \leq k \leq k_{max}$ (where $k_{max}$ is the maximum number of jumps).

The other difference to the deterministic case is that we generally need to apply clustering.
Recall that the transition is typically enabled for several sets in the flowpipe, say, for $j$ (consecutive) sets.
In principle one can apply the reset map to each set individually and spawn $j$ new flowpipes.
But this way the number of flowpipes grows exponentially with the number of jumps.
Instead, approaches in the literature compute an overapproximation of the set union, which is called clustering.
This union of $j$ sets can be represented by several sets again, say $j' < j$, for instance based on an estimated approximation error (see, e.g.,~\cite{FrehseGDCRLRGDM11}), but for simplicity here we just overapproximate the set union with a single set (a zonotope).

\begin{figure}[tb]
	\centering
	\begin{subfigure}[t]{0.48\textwidth}
		\centering
		\includegraphics[width=6cm,height=4cm,keepaspectratio]{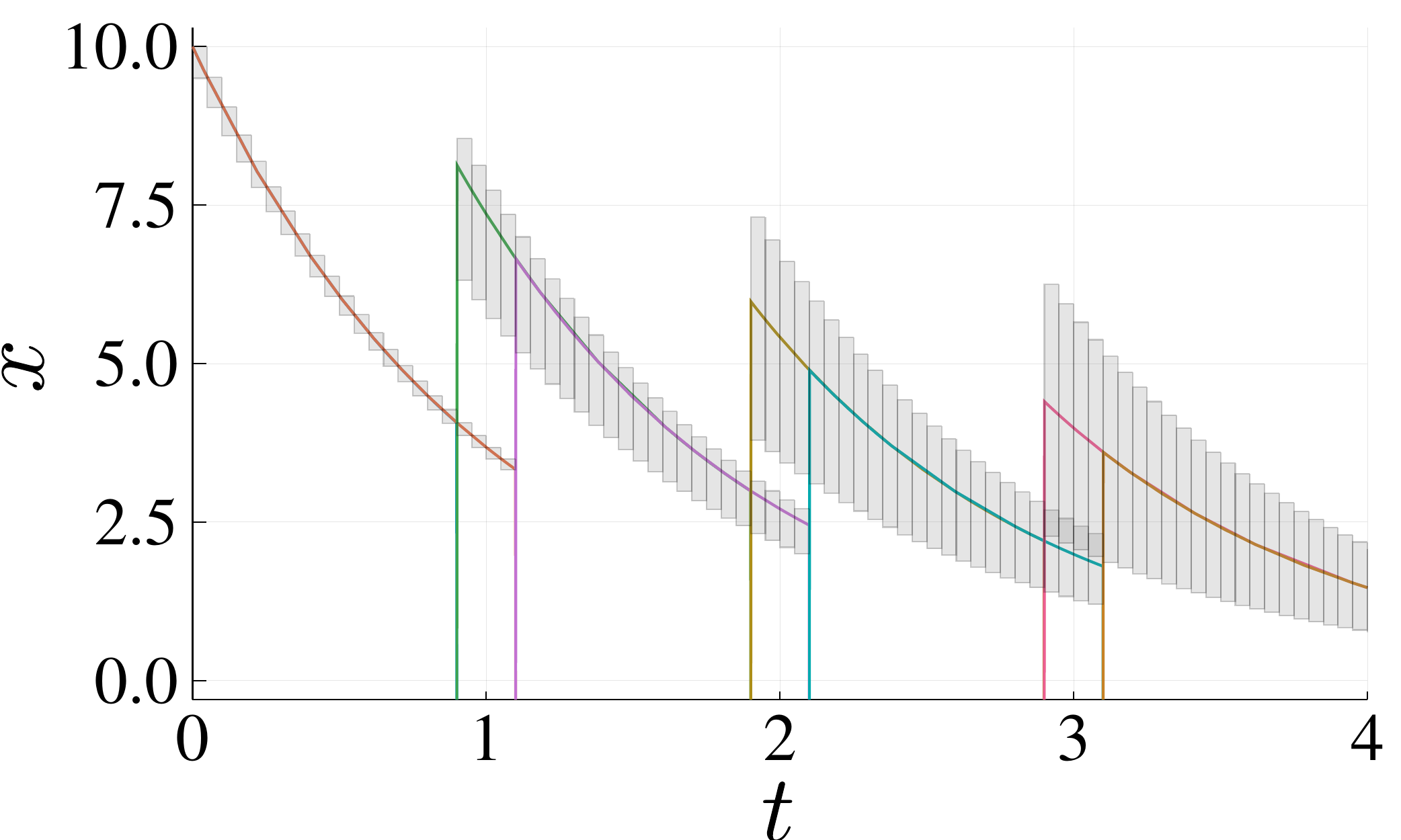}
		\caption{Two trajectories and flowpipes for the nondeterministic instantiation $\jitter = 0.1$ and step size $\delta = 0.05$.}
		\label{fig:simple_example_nondeterministic}
	\end{subfigure}
	\hfill
	\begin{subfigure}[t]{0.48\textwidth}
		\centering
		\includegraphics[width=6cm,height=4cm,keepaspectratio]{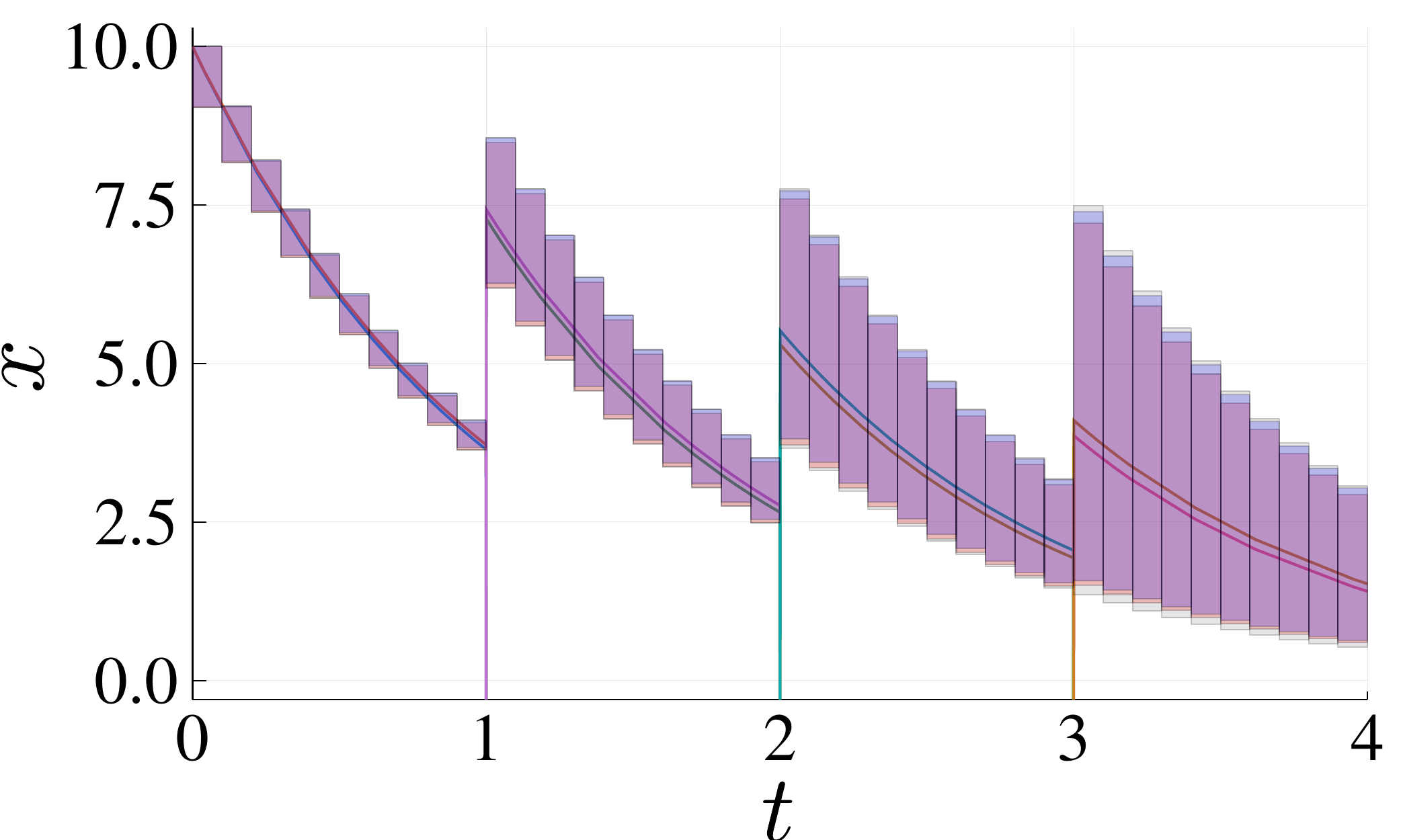}
		\caption{Two trajectories and flowpipes for the parametric setting with interval matrix \A (gray) and with interval matrices $\A_l$ and $\A_h$ (red and blue) and step size $\delta = 0.1$.}
		\label{fig:simple_example_parameters}
	\end{subfigure}
	\caption{Flowpipes for nondeterministic switches and for parameter variation of the model in Figure~\ref{fig:automaton_simple} with $\Tsample = 1$.}
\end{figure}

We illustrate the nondeterministic scenario on the simple model from Figure~\ref{fig:automaton_simple}.
This time we instantiate the model with $\jitter = 0.1$ to make the transition nondeterministic.
The resulting flowpipes are depicted in Figure~\ref{fig:simple_example_nondeterministic}.
We also show the two trajectories that always switch at the earliest (resp.\ latest) point in time.
Observe that, for the rather large time step $\delta = 0.05$ that we chose for illustration purposes, the overapproximation becomes coarser with each jump, which is due to the simple clustering in a single set.
The precision can be improved by choosing a smaller time step.

\subsection{Parametric dynamics}

Now we turn to systems with parametric dynamics, i.e., we consider flow matrices whose entries are intervals.
For a given (possibly interval) matrix $A$ we write $A_{ij}$ for the entry in row~$i$ and column~$j$.
Let \A be an interval matrix.
We say that a scalar matrix $A$ instantiates \A, written $A \in \A$, if they have the same dimension and $A_{ij} \in \A_{ij}$ for every $i$ and $j$.
The semantics of parametric systems is that we consider the whole family of instantiated system.
In other words, given a parametric LTI system $(\A, B, \U)$, we want to compute a flowpipe that covers the behaviors of all (scalar) instantiations $(A, B, \U)$ where $A \in \A$.

We use the algorithm by Althoff et al.~\cite{AlthoffSB07} for computing the flowpipes, which we call the \texttt{ASB07} algorithm.
The algorithm is recursive and hence, unlike in the nonparametric setting, we cannot avoid the wrapping effect~\cite{Althoff10}.
In principle the algorithm can be made non-recursive by iteratively computing (interval) matrix powers, but this problem is known to be NP-hard~\cite{KoshelevaKMN05}, the main issue being that dependencies between the parameter choices get lost in the computations.
We experimented with this idea and mitigation strategies from the literature~\cite{GoldsztejnN14} but overall were not satisfied with the performance.

\medskip

The precision in the parametric setting can be increased in practice by decreasing the interval widths.
Of course, to still consider all allowed behaviors, we would then need to analyze several parametric systems.
We exemplify this idea using a $1 \times 1$ interval matrix $\A := \begin{pmatrix} [-1.01, -0.99] \end{pmatrix}$.
We can split the parameter range into two chunks by considering the matrices $\A_l := \begin{pmatrix} [-1.01, -1] \end{pmatrix}$ and $\A_h := \begin{pmatrix} [-1, -0.99] \end{pmatrix}$.
The semantics outlined above tell us that the behavior of the system using \A is equivalent to the union of the behaviors of systems using $\A_l$ and $\A_h$, respectively.
However, in practice, the union of the latter systems results in more precise approximations because the uncertainty and hence the loss in precision is reduced in each case.
In Figure~\ref{fig:simple_example_parameters} we illustrate the effect on the running example where we replace the original system matrix $\begin{pmatrix} -1 \end{pmatrix}$ by $\A$ respectively $\A_l$ and $\A_h$.
The difference in precision becomes visible in the last flowpipe where the gray parts are not covered by the red and blue flowpipes.

\section{Numerical results} \label{sec:results}

In this section we present results for the electro-mechanical brake model from Section~\ref{sec:model}. The results were obtained on an i7 CPU @ 3.10\,GHz Linux laptop with 16\,GB RAM using Julia v1.4.1~\cite{BezansonEKS17}. The implementation in the library \emph{ReachabilityAnalysis.jl}~\cite{BogomolovFFPS19} and the benchmark scripts are publicly available.\footnote{See \url{http://github.com/JuliaReach/ReachabilityAnalysis.jl}.}

We applied different flowpipe computation techniques and present the results in Section~\ref{ssec:flowpipe}. We report on the analysis of safety properties in Section~\ref{ssec:verification}. Finally, we discuss and compare different verification techniques in Section~\ref{ssec:discussion}.

\subsection{Flowpipe computation} \label{ssec:flowpipe}

\begin{table}[t]
	\centering
	\begin{tabular}{c @{\hspace*{2mm}} c @{\hspace*{2mm}} c @{\hspace*{2mm}} c @{\hspace*{2mm}} c @{\hspace*{2mm}} c @{\hspace*{2mm}} c @{\hspace*{2mm}} c}
		\hline
    \rule{0pt}{4mm} &    & \multicolumn{2}{c}{\textbf{final diameter}} & \raisebox{-1mm}{\textbf{computation}} & \multicolumn{3}{c}{\textbf{\textbf{requirements}}} \\ \cline{3-4}
		\vspace*{1mm}
		\rule{0pt}{4mm} \textbf{$\jitter$ (y/n)} & \textbf{$\delta$ {[}s{]}} & $I$ & $x$ ($\times10^{-5}$) & \textbf{time {[}s{]}} & $\varepsilon$~[m] & $t_c$~[ms] & $v_r$~[mm/s] \\[1mm] \hline
		\rule{0pt}{4mm} \multirow{3}{*}{no}  & $10^{-7}$    & 13.707 & 73.519  & 0.231  & 0.002 & 88.8 & 0.80  \\
		                                     & $10^{-8}$    & 1.369  & 7.343   & 1.08   & 0.002 & 85.8 & 0.81 \\
		                                     & $10^{-9}$    & 0.137  & 0.7343  & 17.0   & 0.002 & 85.5 & 0.81  \\[1mm] \hline

			\rule{0pt}{4mm} \multirow{1}{*}{no (*)}  & $10^{-8}$    & 9.78$\times10^{-6}$ & 0.0000471  & 1.15  & 0.002 & 85.5 & 0.81 \\[1mm] \hline
	
		\rule{0pt}{4mm} \multirow{3}{*}{yes} & $10^{-7}$    & 54.71  & 293     & 0.229  & 0.005 & 64.8 & 1.93 \\
		                                     & $10^{-8}$    & 17.75  & 95.183  & 0.979  & 0.002 & 90.1 & 0.80  \\
		                                     & $10^{-9}$    & 16.56  & 88.8    & 21.1   & 0.01 & 44.7 & 3.84  \\[1mm] \hline
	\end{tabular}
	\vspace*{4mm}
	\caption{Scenarios without parameter variation, using zonotopes of order one in the \texttt{GLGM06} algorithm. Reducing the step size improves the precision. For the cases with jitter, the time interval is $\jitter=[-10^{-8},10^{-7}]$~s.  The requirement coefficients are discussed in Section~\ref{ssec:verification}. The case (*) corresponds to the more precise method described in Section~\ref{ssec:deterministic_switches_improved}.}
	\label{table:GLGM06}
\end{table}

\begin{table}[t]
	\centering
	\begin{tabular}{c @{\hspace*{2mm}} c @{\hspace*{2mm}} c @{\hspace*{2mm}} c @{\hspace*{2mm}} c @{\hspace*{2mm}} c @{\hspace*{2mm}} c @{\hspace*{2mm}} c @{\hspace*{2mm}} c}
		\hline
    \rule{0pt}{4mm} &    &    & \multicolumn{2}{c}{\textbf{final diameter}} & \raisebox{-1mm}{\textbf{computation}} & \multicolumn{3}{c}{\textbf{\textbf{requirements}}} \\ \cline{4-5}
		\vspace*{1mm}
		\rule{0pt}{4mm} \textbf{case} & \textbf{$\jitter$ (y/n)} & \textbf{order} & \textbf{$I$} & \textbf{$x$ ($\times10^{-3}$)} & \textbf{time {[}s{]}} & $\varepsilon$~[m] & $t_c$~[ms] & $v_r$~[mm/s] \\[1mm] \hline
		\rule{0pt}{4mm} \multirow{4}{*}{\emph{pv1}} & \multirow{3}{*}{no} & 1 & 137.25 & 7.305 & 8.817 & 0.005 & 70.5 & 1.89 \\
		&                                                          & 2 & 4.25   & 0.186 & 36.538 & 0.002 & 87.0 & 0.82 \\
		&                                 & 3 & 2.94   & 0.123 & 39.958 & 0.002 & 86.5 & 0.82 \\ \cline{2-9}
		\rule{0pt}{4mm} & yes & 1 & 154.21 & 8.210 & 8.995 & 0.005 & 72.4 & 1.88 \\[1mm] \hline
		\rule{0pt}{4mm} \multirow{4}{*}{\begin{tabular}{@{}c@{}}\emph{pv2}, \\[1mm] $\chi=1\%$\end{tabular}} & \multirow{3}{*}{no}                                                                                                                                                                                                                                                			 & 1 & 2080.79 & 107.708 & 10.63 & $-$ & $-$ & $-$ \\
		&    & 2 & 58.31   & 2.620   & 44.79 & 0.02 & 84.6 & 8.80 \\
		&    & 3 & 39.05   & 1.687   & 45.90 & 0.02 & 58.0 & 8.90 \\ \cline{2-9}
		\rule{0pt}{4mm} & yes & 1 & 2106.50 & 109.84 & 10.24 & $-$ & $-$ & $-$ \\[1mm] \hline
	\end{tabular}
	\vspace*{4mm}
	\caption{Scenarios with parameter variation, and using a fixed step size of $\delta = 10^{-8}$~s in the \texttt{ASB07} algorithm. For the cases with jitter, the time interval is $\jitter=[-10^{-8},10^{-7}]$~s. The requirement coefficients are discussed in Section~\ref{ssec:verification}.
	Dashes ($-$) indicate unsuccessful verification attempts.}
	\label{table:ASB07}
\end{table}

\begin{figure}[tb]
	\centering
	\begin{subfigure}[t]{0.49\textwidth}
		\includegraphics[width=\textwidth,height=8cm,keepaspectratio]{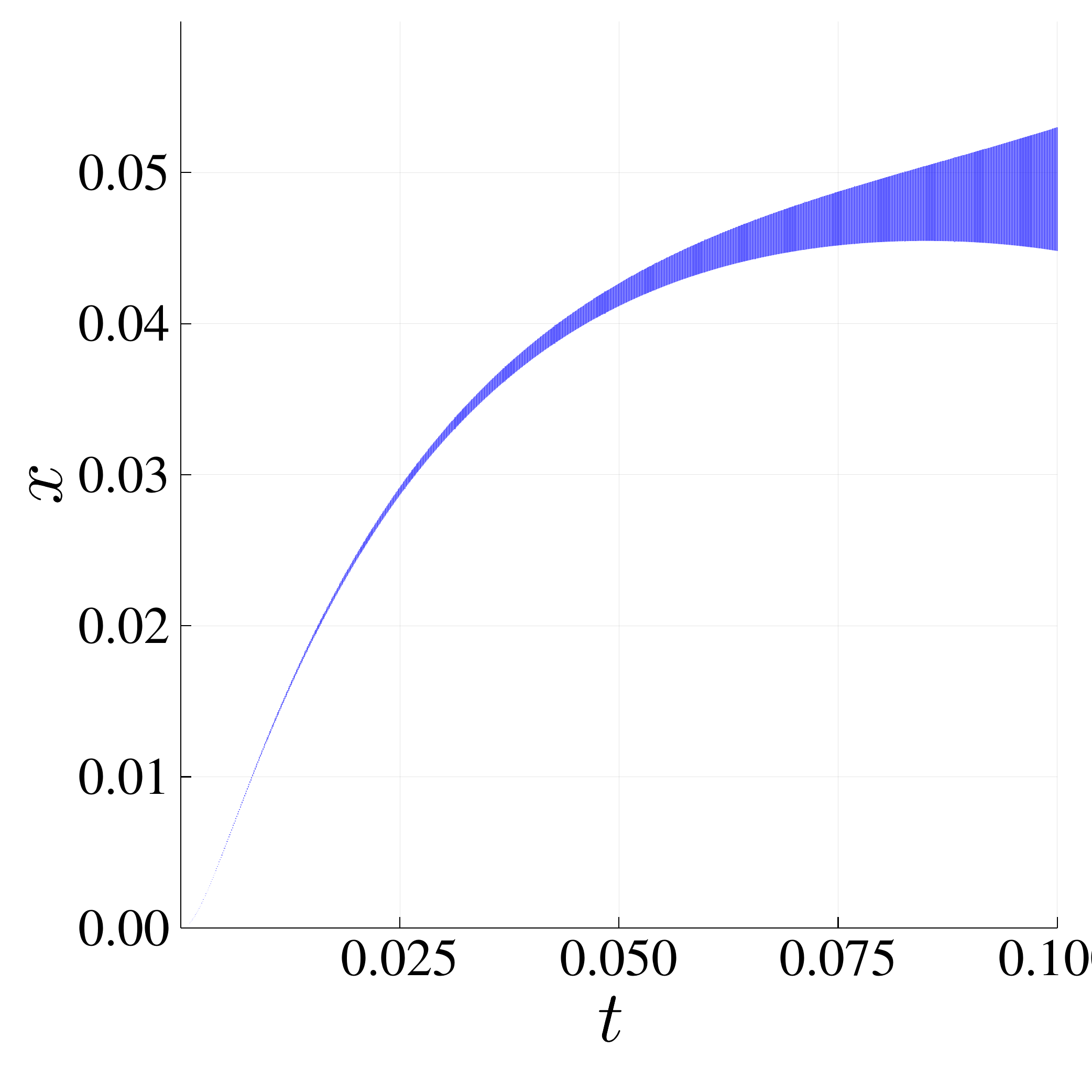}
		\caption{Caliper position $x$.}
		\label{fig:flowpipe_pv1}
	\end{subfigure}
	\hfill
	\begin{subfigure}[t]{0.49\textwidth}
		\includegraphics[width=\textwidth,height=8cm,keepaspectratio]{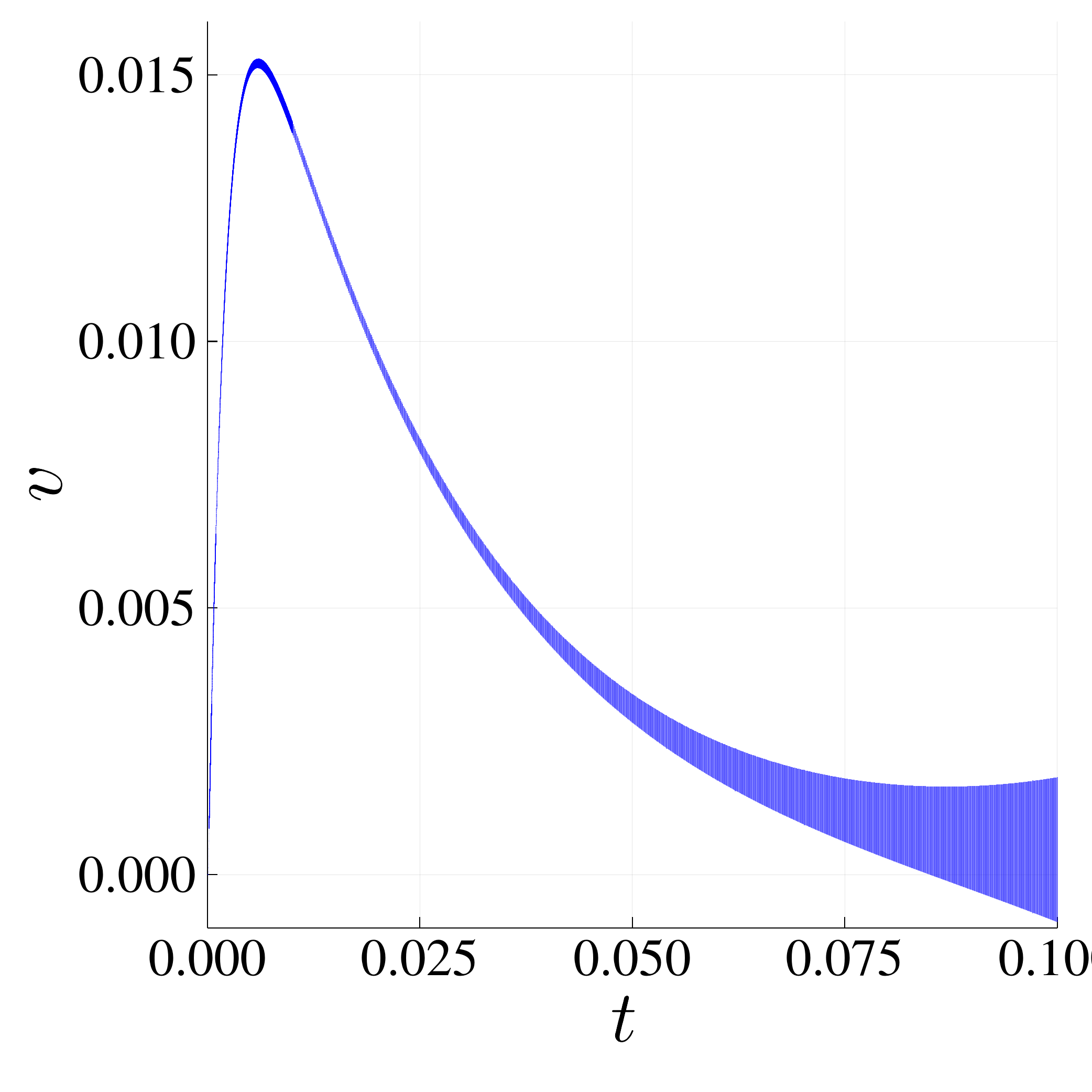}
		\caption{Caliper velocity $v = \dot{x} = \frac{K}{i \cdot d_\mathit{rot}} \cdot I$.}
	\end{subfigure}
	\caption{Flowpipe computation for the case with parameter variation on one coefficient and with jitter (setting~(2)) and a time step $\delta = 10^{-8}$.}
	\label{fig:flowpipes}
\end{figure}

We consider the following settings: (1)~no parameter variation (case \emph{no-pv}), (see Table~\ref{table:GLGM06} for the results); (2)~parameter variation in only one variable (case \emph{pv1}) which combines different physical constants as in~\cite{StrathmannO15}, and (3)~parameter variation in all physical parameters of the model (seven in total; case \emph{pv2}), around 1\% of their nominal value. The results for parameter variation are combined in Table~\ref{table:ASB07} and the reach-set approximation for setting~(2) is plotted in Figure~\ref{fig:flowpipes}. For each of these cases we have considered deterministic switches ($\jitter = 0$) and nondeterministic switches ($\jitter = [-10^{-8}, 10^{-7}]$).
 
In all cases, as a measure of the approximation quality we consider two aspects: the run-time and the width or final diameter of the last flowpipe projected onto variables 1 (electric current $I$) and 2 (position of the caliper $x$). From Table~\ref{table:GLGM06} we see that that reducing the step size $\delta$ improves the precision of the overapproximation error in both variables. It should be noted that in this case, as the system is homogeneous, there is no need to perform zonotope order reduction so we only present results for zonotopes of order one.

In contrast, for the cases with parameter variation (cf.\ Table~\ref{table:ASB07}), we can see that the precision can be increased significantly by taking higher order zonotopes (we use the order-reduction algorithm from~\cite{Girard05}). However, in the presence of jitter, we were not able to improve the precision by using higher-order zonotopes. This would require an accurate overapproximation of the convex hull of a set of zonotopes $\mathit{CH}(Z_1 \cup \cdots \cup Z_k)$ again with a single zonotope, which is out of the scope of this work.

\subsection{Property verification} \label{ssec:verification}

For the algorithms described in this paper, we verify set-based safety properties using a support-function representation. If the flowpipe is represented by zonotopes, support-function evaluations can be performed efficiently~\cite{althoff2016combining}.

There are two requirements for the model of the electro-mechanical brake presented in~\cite{StrathmannO15}: the first requirement refers to a maximum elapsed time $t_c$ since the braking request until the caliper and the disk get in contact. The second requirement is to keep the speed below a certain value $v_r$ upon contact. The stated values for these constants in~\cite{StrathmannO15} are $t_c=23$~ms and $v_r=2$~mm/s. Since the requirements were not verified on the simplified model, neither in~\cite{StrathmannO15} nor in the present work, here we propose a different requirement.

We introduce a tolerance in the caliper position $\varepsilon$ to relax the first condition as: $|x-x_0|\leq\varepsilon$, $\forall t\ge t_c$. Throughout the different scenarios we compute the values for $t_c$ and $v_r$ for certain small values of $\varepsilon$ compared with the disk position $x_0=0.05$. We present the results for \emph{no-pv} and \emph{pv1} in Tables~\ref{table:GLGM06} and~\ref{table:ASB07}, respectively.

In the \emph{no-pv} scenario we verify the requirements for $t_c\approx 90$~ms. Moreover, we found that the caliper speed upon contact is $v_r\approx 0.80$~mm/s. We obtained similar results for the case \emph{pv1} with no jitter. In the presence of jitter, a larger value of $\varepsilon$ was required to obtain similar values of $t_c$ and $v_r$.

\subsection{Comparison with previous work} \label{ssec:discussion}


We note that~\cite{StrathmannO15} does not use our semantics of periodic controllers.
If we compare the run-times presented in Tables~\ref{table:GLGM06} and~\ref{table:ASB07} with those from the original benchmark~\cite{StrathmannO15} we observe two things. First, the run-times for the \empty{pv1} case without jitter in this work ($\approx 9$~s) and in~\cite{StrathmannO15} using the tool Flow* and a different flowpipe computation method ($\approx 48{,}100$~s) differ drastically. In particular, that method does not exploit the linearity of the system and parameters are considered as variables with constant dynamics. This leads to the second observation that modeling parameters as state variables would increase the computational complexity for each parameter, while the algorithm \texttt{ASB07} has the same computational complexity irrespective of the number of parameters.

\section{Conclusions}\label{sec:conclusion}

We have investigated the problem of reachability analysis for periodic time-triggered hybrid systems that model common periodic controllers.
To the best of our knowledge (and surprise), the nondeterministic case of such systems has not been studied before.
We have demonstrated the applicability of our approach on a realistic system of an electro-mechanical brake with a periodic controller.

In the future it would be interesting to study the interplay of our framework with nonlinear models, e.g., the original model from~\cite{StrathmannO15} without simplifications, and extend the framework to algorithms with adaptive step size.


\bibliographystyle{splncs04}
\bibliography{biblio}

\end{document}

%% file: embrake_automaton.tex
\begin{tikzpicture}
	\node[draw,rectangle,rounded corners,thick] (loc) {
		$\begin{array}{@{} c l @{}}
		\dot{I} &= \frac{1}{L} \cdot \left( (K_P \cdot x_e + K_I \cdot x_c) - (R + \frac{K^2}{d_\mathit{rot}}) \cdot I \right) \\[1mm]
		\dot{x} &= \frac{K}{i \cdot d_\mathit{rot}} \cdot I \\[1mm]
		\dot{x}_e &= 0 \\[1mm]
		\dot{x}_c &= 0 \\[1mm]
		\dot{T} &= 1 \\[1mm]
		\hline \\[-2mm]
		\multicolumn{2}{c}{T \leq \Tsample + \jitter}
		\end{array}$};
	\draw[->,>=stealth, loop right,looseness=3] (loc) to node[right] {
		$\begin{array}{@{} c l @{}}
		\multicolumn{2}{c}{T \geq \Tsample - \jitter} \\[1mm]
		\hline \\[-2mm]
		x_e' &:= x_0 - x \\[1mm]
		x_c' &:= x_c + \Tsample \cdot (x_0 - x) \\[1mm]
		T' &:= T - \Tsample
		\end{array}$} (loc);
\end{tikzpicture}

%% file: timeline.tex
\begin{tikzpicture}[scale=0.85,overbrace/.style={decoration={brace},decorate},underbrace/.style={decoration={brace,mirror},decorate}]
	\coordinate (t0) at (0,0);
	\coordinate (t0s) at ($(t0) + (0,-2.1)$);
	\coordinate[right=16mm of t0] (d1);
	\coordinate[right=14mm of d1] (d3);
	\coordinate (d3s) at ($(d3) + (0,-1.3)$);
	\coordinate[right=6mm of d3] (t1);
	\coordinate[right=16mm of d3] (d4);
	\coordinate[right=14mm of d4] (d5);
	\coordinate[right=10mm of d5] (t2);
	\coordinate[right=16mm of d5] (d6);
	\coordinate (d6s) at ($(d6) + (0,-1.3)$);
	\coordinate[right=14mm of d6] (d7);
	\coordinate[right=8mm of d7] (t3);
	\coordinate[right=16mm of d7] (d8);
	\coordinate (d8s) at ($(d8) + (0,-2.1)$);
	\draw (t0) -- (d1);
	\draw[dotted] (d1) -- (d3);
	\draw (d3) -- (d4);
	\draw[dotted] (d4) -- (d5);
	\draw (d5) -- (d6);
	\draw[dotted] (d6) -- (d7);
	\draw (d7) -- (d8);
	\draw[very thick] (t0) node[above=2mm] (t0lab) {$t_0$} +(0,0.2) -- +(0,-0.2);
	\draw (d1) -- +(0,0.2) -- +(0,-0.2) node[below] {$t_0 + \delta$};
	\draw (d3) -- +(0,0.2) -- +(0,-0.2) node[below] {$t_0 + k_1 \delta$};
	\draw[very thick,color=red] (t1) node[above=2mm] (t1lab) {$t_1$} +(0,0.2) -- +(0,-0.2);
	\draw (d4) -- +(0,0.2) -- +(0,-0.2) node[below=4mm] {$t_0 + (k_1 + 1) \delta$};
	\draw (d5) -- +(0,0.2) -- +(0,-0.2) node[below] {$t_0 + (k_2 - 1) \delta$};
	\draw[very thick,color=blue] (t2) node[above=2mm] (t2lab) {$t_2$} +(0,0.2) -- +(0,-0.2);
	\draw (d6) -- +(0,0.2) -- +(0,-0.2) node[below=4mm] {$t_0 + k_2 \delta$};
	\draw (d7) -- +(0,0.2) -- +(0,-0.2) node[below] {$t_0 + (k_3 - 1) \delta$};
	\draw[very thick,color=green!70!black] (t3) node[above=2mm] (t3lab) {$t_3$} +(0,0.2) -- +(0,-0.2);
	\draw (d8) -- +(0,0.2) -- +(0,-0.2) node[below=4mm] {$t_0 + k_3 \delta$};
	\draw[underbrace] (d3s.south) -- node[below=1mm] {computed transition time frame} (d6s.south);
	\draw[overbrace] (t1lab.north) -- node[above=1mm] {actual transition time frame} (t2lab.north);
	\draw[underbrace] (t0s.south) -- node[below=1mm] {computed flowpipe time frame} (d8s.south);
	\draw[overbrace] ($(t0lab.north) + (0,0.7)$) -- node[above=1mm] {actual flowpipe time frame} ($(t3lab.north) + (0,0.7)$);
\end{tikzpicture}

%% file: simple_automaton.tex
\begin{tikzpicture}
	\node[draw,rectangle,rounded corners,thick] (loc) {
		\begin{tabular}{@{} c @{}}
		$\dot{x} = -x$ \\[1mm]
		$\dot{T} = 1$ \\[1mm]
		\hline \\[-2mm]
		$T \leq \Tsample + \jitter$
		\end{tabular}};
	\draw[->,>=stealth, loop right,looseness=3] (loc) to node[right] {
		\begin{tabular}{@{} c @{}}
		$T \geq \Tsample - \jitter$ \\[1mm]
		\hline \\[-2mm]
		$x' := 2 x$ \\[1mm]
		$T' := T - \Tsample$
		\end{tabular}} (loc);
\end{tikzpicture}